\documentclass[12pt]{article}
\usepackage{setspace}
\usepackage{graphicx}
\usepackage{epstopdf} 
\title{Direct optical nanoscopy with axially localized detection}

\usepackage[T1]{fontenc}
\usepackage[utf8]{inputenc}
\usepackage{authblk}
\usepackage{amsmath,amsfonts,amssymb}
\usepackage[squaren,Gray]{SIunits}
\usepackage{color}
\usepackage{ulem}

\author[1,2]{N. Bourg}
\author[2]{C. Mayet}
\author[2]{G. Dupuis}
\author[3]{T. Barroca}
\author[3]{P. Bon}
\author[2]{S. L\'{e}cart}
\author[3]{E. Fort}
\author[1,2,*]{S. L\'{e}v\^{e}que-Fort}
\affil[1]{Institut des Sciences Mol\'{e}culaires d'Orsay (ISMO), Universit\'{e} Paris-Sud, CNRS UMR 8214, F91405 Orsay Cedex, France}
\affil[2]{Universit\'{e} Paris-Sud, Centre de Photonique BioM\'{e}dicale (CPBM), F\'{e}d\'{e}ration LUMAT, CNRS FR 2764, F91405 Orsay Cedex, France}
\affil[3]{Institut Langevin, ESPCI ParisTech, CNRS, PSL Research University, 1 rue Jussieu, F-75005 Paris, France}
\affil[*]{Corresponding author: sandrine.leveque-fort@u-psud.fr}

\begin{document}
\begin{spacing}{1.5}

\maketitle

\newpage
\noindent\textbf{Evanescent light excitation is widely used in super-resolution fluorescence microscopy to confine light and reduce background noise. Herein we propose a method of exploiting evanescent light in the context of emission. When a fluorophore is located in close proximity to a medium with a higher refractive index, its near-field component is converted into light that propagates beyond the critical angle. This so-called Supercritical-Angle Fluorescence (SAF) can be captured using a high-NA objective and used to determine the axial position of the fluorophore with nanometer precision. We introduce a new technique for 3D nanoscopy that combines direct STochastic Optical Reconstruction Microscopy (dSTORM) imaging with dedicated detection of SAF emission. We demonstrate that our approach of a Direct Optical Nanoscopy with Axially Localized Detection (DONALD) yields a typical isotropic 3D localization precision of 20-nm.} \\

\newpage
\noindent Super-localization methods in optical microscopy such as PALM, (F)PALM and dSTORM, have shattered the spatial resolution barrier imposed by the diffraction limit. These techniques are based on the sequential detection of several thousands individual fluorescent molecules \cite{Betzig2006,Hess2006,MichaelJRustMarkBates2006,VandeLinde2011}. In their simplest implementations, these methods typically improve lateral resolution by one order of magnitude, however their axial resolution is still limited by diffraction. Specific techniques must be developped to tackle this strong anisotropy resolution, which compromises 3D imaging. Super-localization techniques must be combined with Point Spread Function (PSF) engineering methods, to measure the depth position of each detected fluorophore \cite{Huang2008,Dahan2012,XuKe;BabcockHazenP;Zhuang2012,Rama2009,Badieirostami2010}.  Single- \cite{Huang2008} and double- \cite{XuKe;BabcockHazenP;Zhuang2012} cylindrical-lens methods have achieved axial-localization precisions 60 and 20~nm, respectively. The former is very stable and straightforward to implement, whereas the higher precision of the latter comes at the cost of increased complexity. Similarly, alternative elaborate methods such as interferometric PALM (iPALM) \cite{Shtengel2008} and the Self-Bending PSF (SB-PSF) \cite{Jia2014} can achieve axial localization precision of 10 nm and 15~nm, respectively\cite{Klein2014}. In addition, all these strategies provide only the relative axial positions of the fluorophores with respect to an arbitrary focal plane. Hence, 3D optical nanoscopy is in need of a method that combines high nanometer axial precision, simplicity of implementation and absolute axial positioning. \\

\noindent Herein we report a new approach termed “Direct Optical Nanoscopy with Axially Localized Detection” (DONALD). This type of nanoscopy combines standard super-localization technique with Supercritical-Angle Fluorescence (SAF) analysis \cite{Ruckstuhl2000a}. The latter is based on the light emission above the critical angle that occurs when fluorophores are placed in the vicinity of the coverslip interface. Within this region, DONALD achieves an iso-3D nanometer resolution, yielding the absolute axial position of the fluorophores with a precision of 10 to 20 nm. \\

\noindent  A fluorophore can be modeled as a dipolar emitter radiating in the far field. This dipole is also endowed with a non propagative near field component that depends on the surrounding refractive index $n_m$. In the presence of an interface with a medium with a refractive index of $n_g>n_m$, the transmitted light follows Snell-Decartes law of refraction (Fig. 1a). The refracted light is emitted within a cone that is limited by the critical angle $\theta_c = \arcsin{(n_{m} / n_{g})}$. This component is referred to as Under-critical Angle Fluorescence (UAF). However, if the fluorophore-interface distance $d$ is smaller than the fluorescence wavelength $\lambda_{em}$, then additional SAF emission is observed.  The evanescent near field component in the homogeneous medium surrounding the fluorophores, with an index of refraction of $n_m$, becomes propagative beyond the critical angle $\theta_c$ inside the medium of higher refractive index, $n_g>n_m$. SAF emission can be detected for fluorophores in the cellular medium located in the vicinity of the coverslip. The SAF intensity is potentially equal to as much as 50$\%$ of all fluorescence emitted into the coverslip when the fluorophore is in direct contact with the interface ($d = 0$) (Fig. 1b)\cite{Fort2008}. \\

\noindent Whereas the number of UAF photons $N^{UAF}$ remains nearly constant as a function of the interface-fluorophore distance $d$, the number of SAF photons $N^{SAF}$ decreases approximately exponentially \cite{Ruckstuhl2003}.  Hence, the simultaneous measurement of $N^{SAF}$ and $N^{UAF}$ and the computation of the fluorophore SAF ratio $\rho^{SAF} =N^{SAF}/N^{UAF}$ for each detected fluorophore can be used to determine the absolute axial position of the fluorophore, $d$. Here, we propose that this principle may be used to achieve axial localization with a nanometer precision. Ruckstuhl et al. have already successfully applied this principle using a home-made parabolic objective \cite{Ruckstuhl2003,Winterflood}. Unfortunately, this point scanning objective is not compatible with lateral super-localization. Current commercial high-numerical-aperture objectives ($NA \sim n_{g} > n_m$) allow the efficient collection of SAF emission \cite{Barroca2011}. The angular distributions of both the SAF and UAF components can be directly observed in their Back-Focal Plane (BFP). These aplanetic objectives satisfy Abbe sine relation: light emitted with an angle $\theta$ lies within a circle of radius $\rho=n f \sin(\theta)$ in the BFP where $n$ is the refractive index of the immersion medium and $f$ is the focal length of the objective. Hence, the UAF emission lies within a disk of radius $n f \sin(\theta_c)$, and the SAF component has a ring shape and surround the UAF disk up to a radius of $f NA$ (Fig.~1b). \\

\noindent Various strategies can be implemented to discriminate between the UAF and SAF components in the BFP.  A straightforward method of selecting for SAF emission consists of using a disk-shaped mask in the BFP to block the UAF emission. The diameter of the resulting PSF, $\sigma_{PSF}^{SAF}$, is 1.7 times larger than that of the standard PSF \cite{Barroca2011}, and $N^{SAF}$ represents less than 50$\%$ of the total number of collected photons, $N^{EPI}$. However, this strategy causes degradation of the lateral 2D localization precision $\sigma$ because in STORM-like microscopes, $\sigma \propto {\sigma_{PSF}}/{\sqrt{N}}$ \cite{Thompson2002}. The detection of SAF emission for enhanced axial localization should not come at the cost of a degradation in lateral resolution. Our alternative approach extracts $N_{SAF}$ by measuring $N_{EPI}$ and $N_{UAF}$ on two simultaneously acquired PSFs. Both PSFs are well defined, with a significant number of photons, and can thus be used to compute the lateral 2D super-localization with a good signal-to-noise ratio. By simultaneously capturing $N^{EPI}$ and $N^{UAF}$, we can determine $N^{SAF} = N^{EPI}-N^{UAF}$. Notably, this approach is the single-molecule analog of the full-field virtual SAF technique \cite{Barroca2012,Barroca2013}. Finally, by computing $\rho^{SAF}$, we achieve axial super-localization of the fluorophore. \\

\noindent Our experimental setup consists of a home-made DONALD module that we inserted between the output of a standard full-field microscope and an EMCCD camera. This module uses a beamsplitter to split the fluorescence emission into two imaging paths (Fig. 2a). The first EPI path is directly imaged on half of the EMCCD detector and is used to compute $N^{EPI}$ for a given PSF. On the second path, the SAF ring is blocked out in the image plane of the objective BFP in order to generate a corresponding UAF-only PSF on the other half of the EMCCD detector. $N^{UAF}$ is computed from this image. When a fluorophore is imaged on both paths, it is first super-localized in 2D using a wavelet segmentation algorithm \cite{Izeddin2012}. Then, we measure $N^{EPI}$\ and $N^{UAF}$ via numerical integration within a PSF region of $9\times9$ pixels. $\rho^{SAF}(d)$ is then computed to determine the depth $d$ of the fluorophore (Fig. 2b; see also Supplementary Fig. 1). The precision of axial localization depends on $n_m$, $n_g$ and the Signal-to-Noise Ratio ($SNR$) which is defined as  $SNR = I_{max}^{EPI}/\sqrt{2\pi I_{max}^{EPI}}$,  where $I_{max}^{EPI}$ is the maximum intensity of the EPI PSF \cite{Izeddin2012}. We performed Monte Carlo simulations of $\rho^{SAF}$ as a function of $d$ (with $0 < d < \lambda_{em}$) for our typical experimental conditions ($n_m = 1.33$, $n_g = 1.515$, $SNR = 5.3$ or 7.4) \cite{Tang2007} (see Fig.~3a; see also Supplementary Fig. 2). We then determine the axial localization precision which  decreases with increasing axial position of the fluorophore $d$ (Fig.~3b; see Supplementary Fig. 2 for more details regarding the simulations). For a typical $SNR = 7.4$, we found that the axial localization precision is better than 20 nm when $0 < d < 0.2\lambda_{em}$ and better than 40 nm when $0.2\lambda_{em} < d < 0.5\lambda_{em}$. To confirm the results of these simulations, we performed experimental calibrations (See Supplementary methods). We used 20-nm fluorescent nanospheres embedded in a 3\%-agarose gel  ($n_{m} = n_{water} = 1.33$) deposited on a coverslip ($n_g = 1.515$). For each bead, we measured both the $\rho^{SAF}$ with the DONALD module and the depth $d$ using the cylindrical lens method, which determine the axial position d from the PSF shape \cite{Huang2008}. The experimental results and the theoretical calculations were in excellent agreement (see the black circled red dots and black line, respectively, in Fig.~3a). The theory can thus be used directly to convert measured $\rho^{SAF}$ value into the axial position of the fluorophore.\\

\noindent We demonstrated the performance of DONALD in cell imaging. We first imaged dSTORM frames of F-actin labeled with phalloidin-Alexa 488 of COS-7 cells immersed in a thiol+oxygen scavenger buffer (See Supplementary methods) with a refractive index of $n_m = 1.33$\cite{VandeLinde2011} (Fig.~4a \& 4b). Fluorophore excitation was achieved using a blue laser ($\lambda_{em} = 488~nm$) in the TIRF configuration. The laser power was held constant throughout the entire imaging process ($2~kW \cdot cm^{-2}$). The number of detected molecules in each frame remained stable throughout the entire acquisition (15,000 frames) by virtue of the amino acid residue tryptophan, which is a component of the actin-probe phalloidin. Tryptophan quenches rhodamine dyes such as Alexa Fluor 488\cite{Nanguneri2014}. Using the DONALD technique, we achieved axial resolution of $35 \pm 1$~nm for structures located approximately 105~nm away from the coverslip (Fig.~4c, 4d, 4e: Zones 1 \& 3). For filaments located at $d \approx 180$~nm and $d \approx 148$~nm, we measured axial resolution of 62~nm and 54~nm, respectively (Fig.~4b, 4d, 4e: Zones 2 \& 4). As expected from the model, the closer to the coverslip these features were located, the better was the axial resolution.
We also imaged a microtubule network of CHO cells that been labeled with antibody-Alexa647 and were immersed in a buffer composed of 50$\%$ PBS and 50$\%$ Vectashield\cite{Olivier2013a} (Fig.~5a, 5b). We used ViSP-software\cite{ElBeheiry2013} to create a three-dimensional projection of a sub-region of the DONALD image (Fig.~5b: white box, 5c), on which we plotted a XZ projection plane of two microtubules (Fig.~5c: white box, 5d). We measured a difference of these two microtubules axial positions of aproximately 50-60~nm. In addition, an estimated value of 65~nm was obtained for the diameter of the microtubule based on axial positioning measurements of a microtubule located approximately 130~nm from the coverslip (Fig.~5b, 5c, 5e: Zone 5). The microtubule diameter (20-30~nm) is typically measured to be approximately 50-60~nm because of the additional size of the primary and secondary antibodies\cite{Jia2014}. There are two reasons for slightly higher diameter measurement. First, the higher refractive index of the Vectashield-based buffer ($n_m = 1.38$ compared with 1.33 for the thiol-based buffer) causes the critical angle to be higher, thus lowering the SAF ratio for a given objective. Second, in our case, the long-term efficiency of the Vectashield-based buffer was not as good as in the case of the thiol-based buffer, hence limiting the number of frames that could be acquired with an acceptable detection density (typically 2500 vs. 15,000). Regardless of these drawbacks, because Alexa Fluor 647 has both a higher quantum yield and a higher emission wavelength than does Alexa Fluor 488, it allows a deeper imaging depth while providing a nanometer axial resolution comparable to that achievable using PSF engineering techniques (Fig.~5b, 5e: Zone 6, $d \approx 280$~nm, $FWHM = 100$~nm).\\

\noindent In summary, DONALD is a new 3D super-localization nanoscopy technique based on detection of the near-field supercritical emission (SAF). The achieved precision of axial localization reaches 20 nm within the nearest 150~nm to the coverslip, thus offering isotropic localization for a typical dSTORM coupling. For distances between 180 nm and 500 nm from the coverslip, the DONALD resolution is at least equivalent to that achieved using current PSF engineering methods (approximately 60~nm). However, a unique feature offered by DONALD is access to the absolute depth of the fluorophore with respect to the coverslip. DONALD is not sensittive to local thickness variations and/or the tilt of the coverslip. In addition, whereas PSF engineering techniques are limited to a low concentration of fluorophores because of the enlargement of the PSF volume and hence to tend to require the use of a high laser power, DONALD can maintain the imaging speed of a standard 2D super-localization microscope. According to our simulations, an axial localization precision of 10 to 5 nm souhld be achievable and improvements in DONALD localization efficiency are limited by the SNR. Further improvement could be achieved by using brighter dyes and/or alternative buffers. In a complementary manner, higher-NA objectives with NAs of 1.65 and 1.70 could allow an increase in the SNR through enhancement of the SAF emission, thereby further improving the axial localization precision. \\

\noindent \textbf{Methods}

\noindent \textbf{Optical set-up}. 3D super-localization images were acquired using a Nikon Eclipse Ti inverted microscope combined with a Perfect Focus System and configured for these studies in TIRF excitation. Samples were excited with 488-nm (Genesis MX-STM 500 mW, Coherent) and 637-nm (Obis 637 LX 140 mW, Coherent) optically pumped semiconductor laser. A set of full-multiband laser filters, optimized for 405-, 488-, 561- and 635-nm laser sources (LF405/488/561/635-A-000, Semrock) was used to excite Alexa Fluor 488 or 647 for the collection of the resultant fluorescence via a Nikon APO TIRF 60x, NA 1.49 oil immersion objective lens. All images were recorded using a 512x512 pixels EMCCD camera (iXon 897, Andor), split on two regions of 256x256-pixel area and positioned on the focal plane of the DONALD module (2.7x magnification, optical pixel size of $\sim$100 nm). \\

\noindent \textbf{dSTORM imaging}. To induce the majority of the fluorophores into the dark state, we excited the samples using a laser in an oblique configuration (488 nm for F-Actin labeled with Alexa-488 and 647 nm for microtubules immunolabeled with Alexa-647). Once the density of fluorescent dye was sufficient (typically, < 1 molecule/$\mu$m$^2$), we switched on the laser that was used in TIRF excitation with an irradiance of 2 kW/cm-2 and activated the real-time three-dimensional localization performed by home-written Python code (See supplementary Fig. 1 for more details regarding the 3D localization). For all recorded images, the integration time and the EMCCD gain was set to 50 ms and 150, respectively. \\

\noindent \textbf{Acknowledgements} \\
The authors thank J. Dompierre for fruitful help on the immunofluorescence.
We acknowledge the financial support of the AXA Research Fund and the French National Research Agency (Project SMARTVIEW). \\

\noindent \textbf{Author contributions} \\
N.B., G.D., E.F. and S.L.F. conceived and designed the project. N.B. performed the experiments, simulations and analysis. C.M. and N.B. developed the photoswitching buffer. C.M., N.B. and S.L. optimized the immunofluorescence protocol. T.B. and P.B. helped on the simulation and the DONALD module. All authors wrote the manuscript.  \\

\noindent \textbf{Additional information} \\
Supplementary information is available in the online version of the paper. Correspondence and requests for materials should be adressed to S.L.F. \\

\noindent \textbf{Competing financial interests} \\
The authors declare no competing financial interests. \\

\begin{figure}
\includegraphics[scale=0.805]{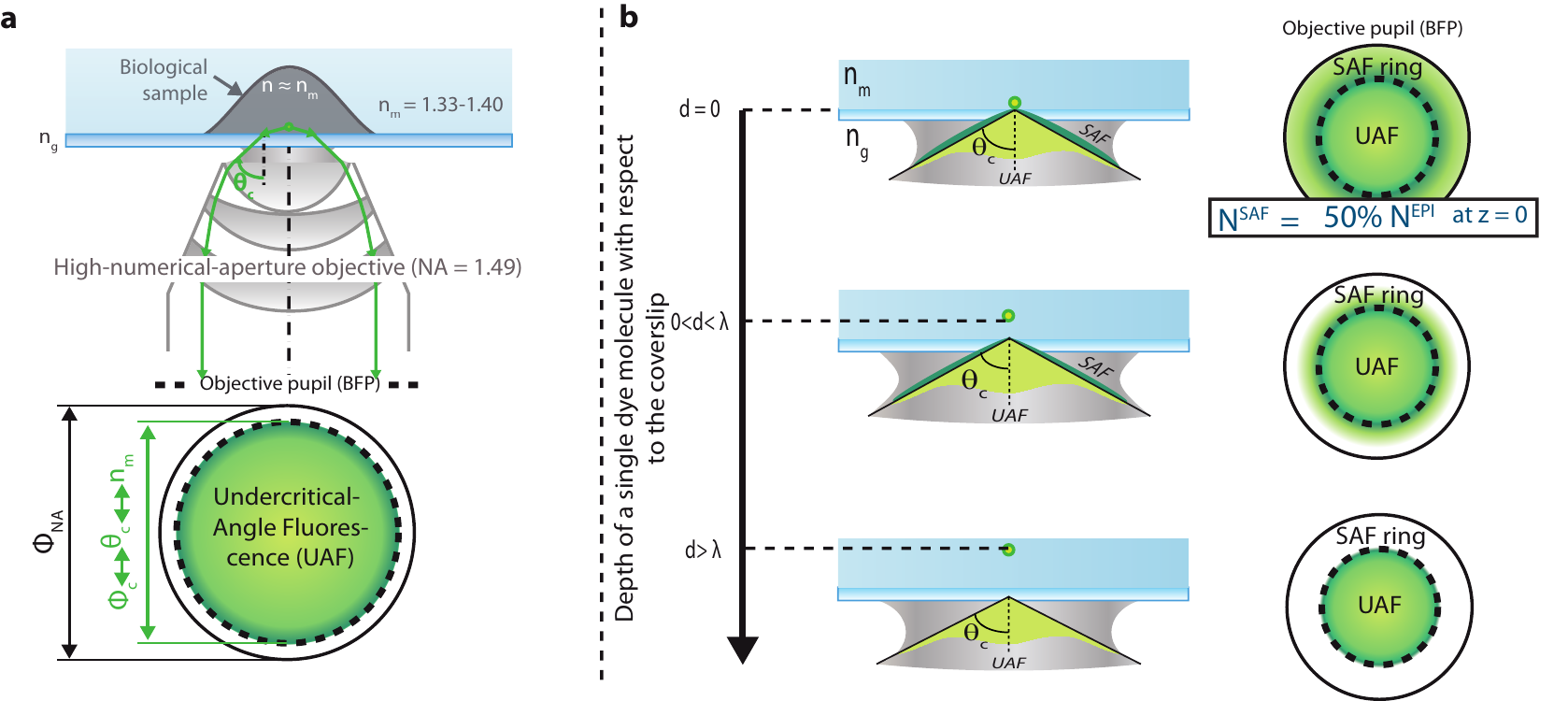} 
\caption{\textbf{Far and near field emission components. a} The far-field emission component (Undercritical-Angle Fluorescence, UAF)  has an angular distribution determined by the law of refraction and limited by the critical angle $\theta_c$. This angular distribution of light can be retrieved on the Back Focal Plane (BFP) within a plane disk of diameter $\phi_c$, which is related to $\theta_c$. \textbf{b} A porition of the near field component, known as Supercritical-Angle Fluorescence (SAF), of a dye molecule located in the near-field region (0 to $\lambda_{em}$) is collected by the objective beyond $\theta_c$. The number of SAF photons, $N^{SAF}$, is potentially equal to as much as 50\% of $N^{EPI}$ when the dye is in close proximity to the coverslip and decreases exponentially as the dye depth d increases.}  
\end{figure}

\begin{figure}
\includegraphics[scale=0.75]{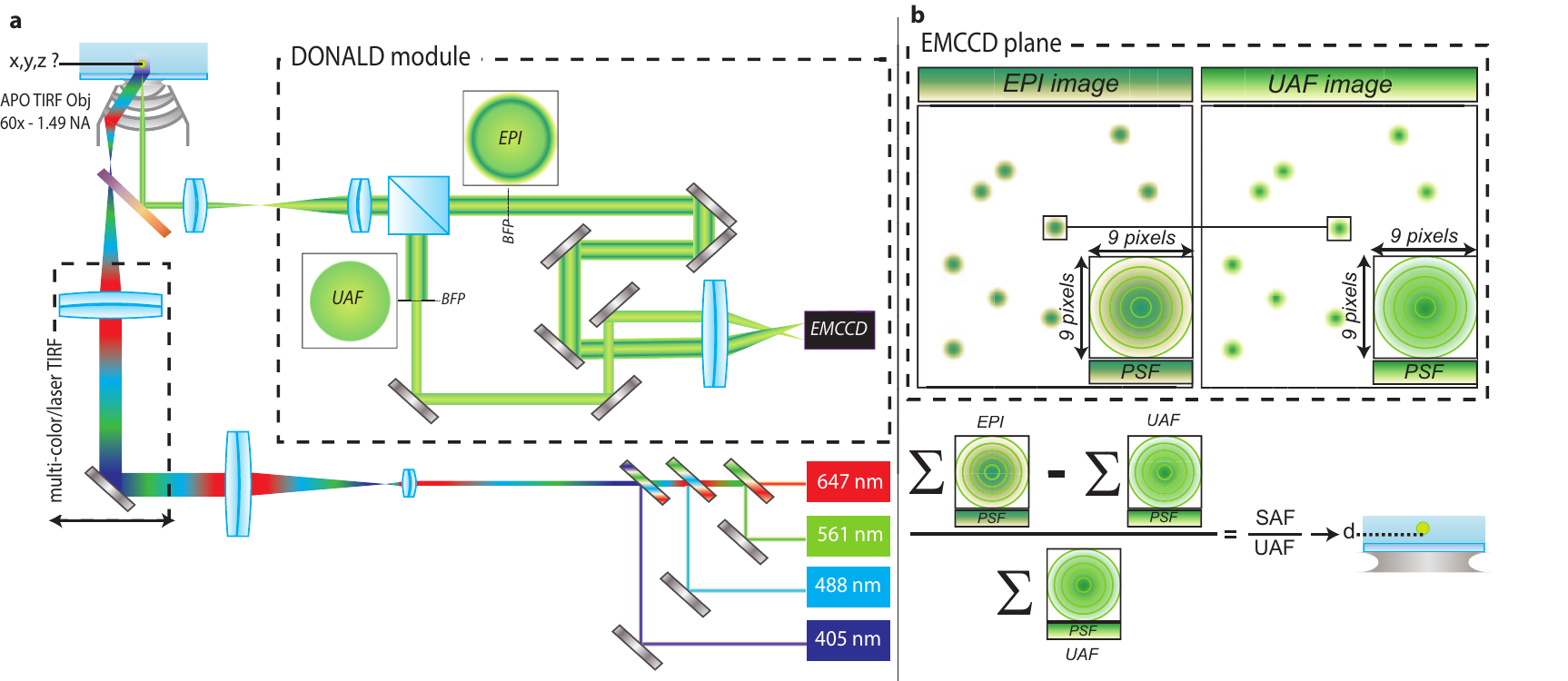} 
\caption{\textbf{Direct Optical Nanoscopy with Axially Localized Detection. a} Schematic diagram of the experimental set-up: A homemade multi-color/laser TIRF stage is connected to the input of a conventional wide-field microscope. TIRF excitation light passes through a 4 colors filter-set (405, 488, 561, 647) and an apochromatic TIRF objective of NA = 1.49. The fluorescence emission of dye molecules is collected by the objective and reflected to the DONALD module which splits the fluorescence into two parts: the EPI part is directly imaged on one half of an EMCCD, and the UAF part (the SAF component is blocked in the BFP) is recorded on the other half. \textbf{b} DONALD data analysis: First, in the UAF and EPI portions of the frame, each PSF is super-localized in 2D and the number of UAF and EPI photons ($N^{UAF}$ and $N^{EPI}$, respectively) are calculated via signal integration in a 9x9 area of pixels. Finally, the SAF ratio is computed and converted into the absolute dye depth d.} 
\end{figure}

\begin{figure}
\includegraphics[scale=1]{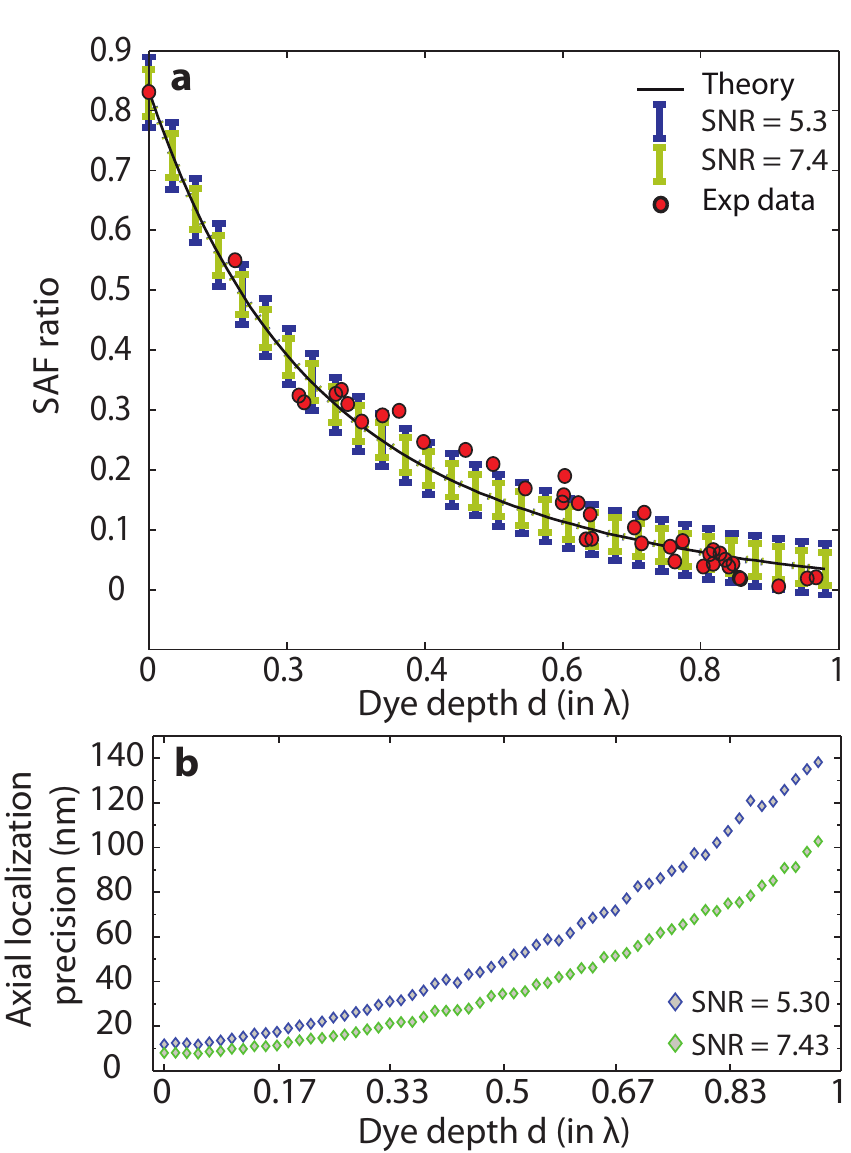} 
\caption{\textbf{DONALD theory. a} Theoretical decay of the SAF ratio (black line) as a function of the dye depth d for $n_m = 1.33$ and $n_g = 1.515$. 5000 iterations of a Monte-Carlo simulation of this theory were performed for 2 different SNRs : 5.3 (blue error bars) and 7.4 (green error bars). \textbf{b} The Monte-Carlo simulations were used to calculate the axial localization precision for both SNRs. Experimental verification of the theory was performed using 20 nm red-beads embedded in 3\% agarose gel ($n_m = 1.33$). The DONALD module was used to measure the SAF ratio, and a PSF shaping method (cylindrical lens) was applied to determine the depth. The experimental results (\textbf{a}, black circled red dot) are consistent with and confirm the DONALD theory.} 
\end{figure}

\begin{figure}
\includegraphics[scale=0.55]{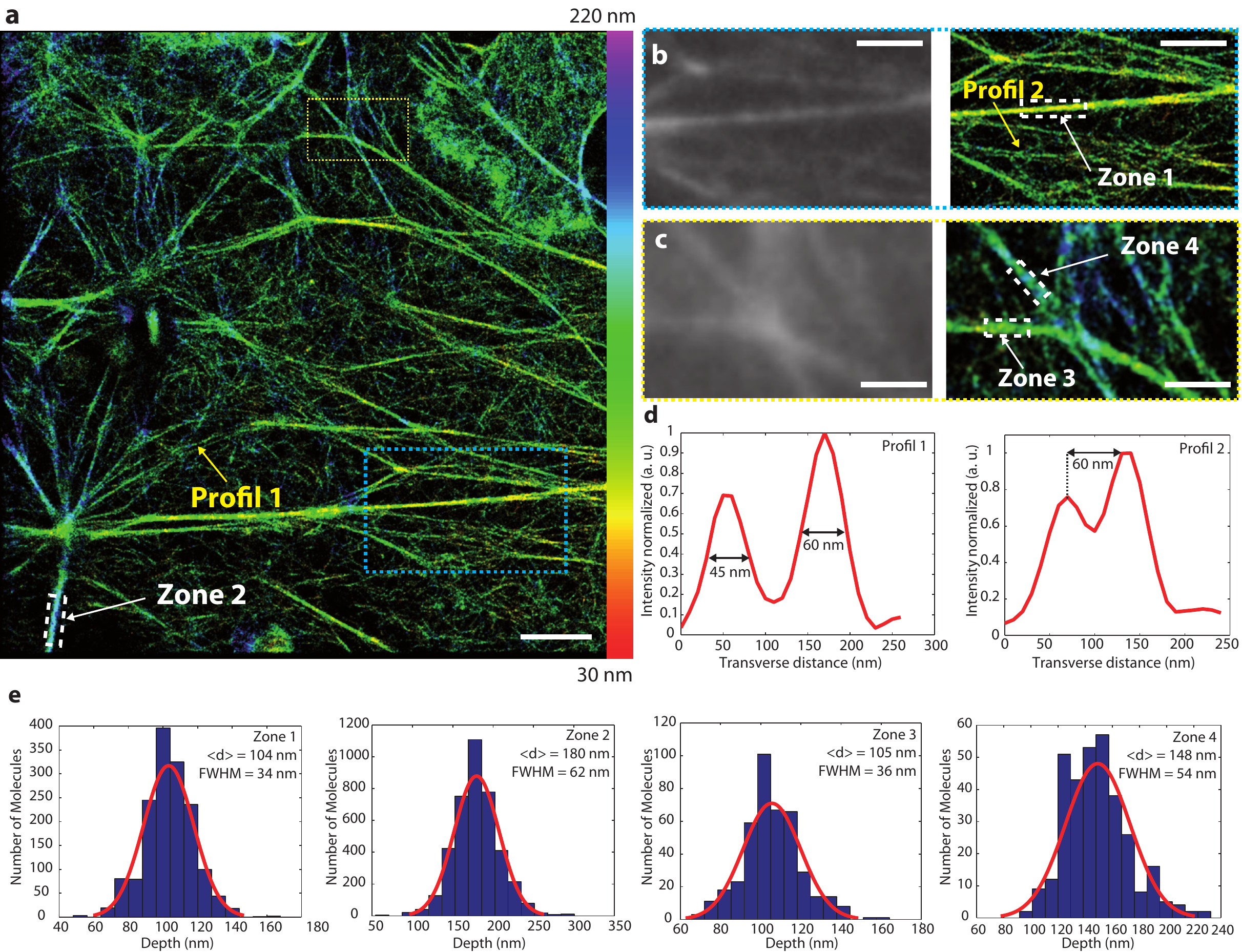} 
\caption{\textbf{dSTORM imaging of F-actin in COS-7 cells immersed in a thiol+oxygen scavenger buffer using DONALD. a} Three-dimensional DONALD image in which the depth is color-coded as indicated by the colored depth-scale bar. \textbf{b, c} Conventional (left) and three-dimensional (right) images of sub-areas of (\textbf{a}), as indicated by the blue and yellow boxes. \textbf{d} Transverse profiles of "Profil 1" and "Profil 2" in (\textbf{a}) and (\textbf{b}). \textbf{e} Axial profiles of various structures (Zones 1-4) in (\textbf{a}) and (\textbf{b}). Axial resolution of 34-36 nm, 54 nm and 62 nm were achieved for filaments located at depths of  105 nm, 148 nm and 180 nm, respectively. Scale bars: 3 $\mu m$ (\textbf{a}), 2 $\mu m$ (\textbf{b}), 1 $\mu m$ (\textbf{c}).} 
\end{figure}

\begin{figure}
\includegraphics[scale=0.45]{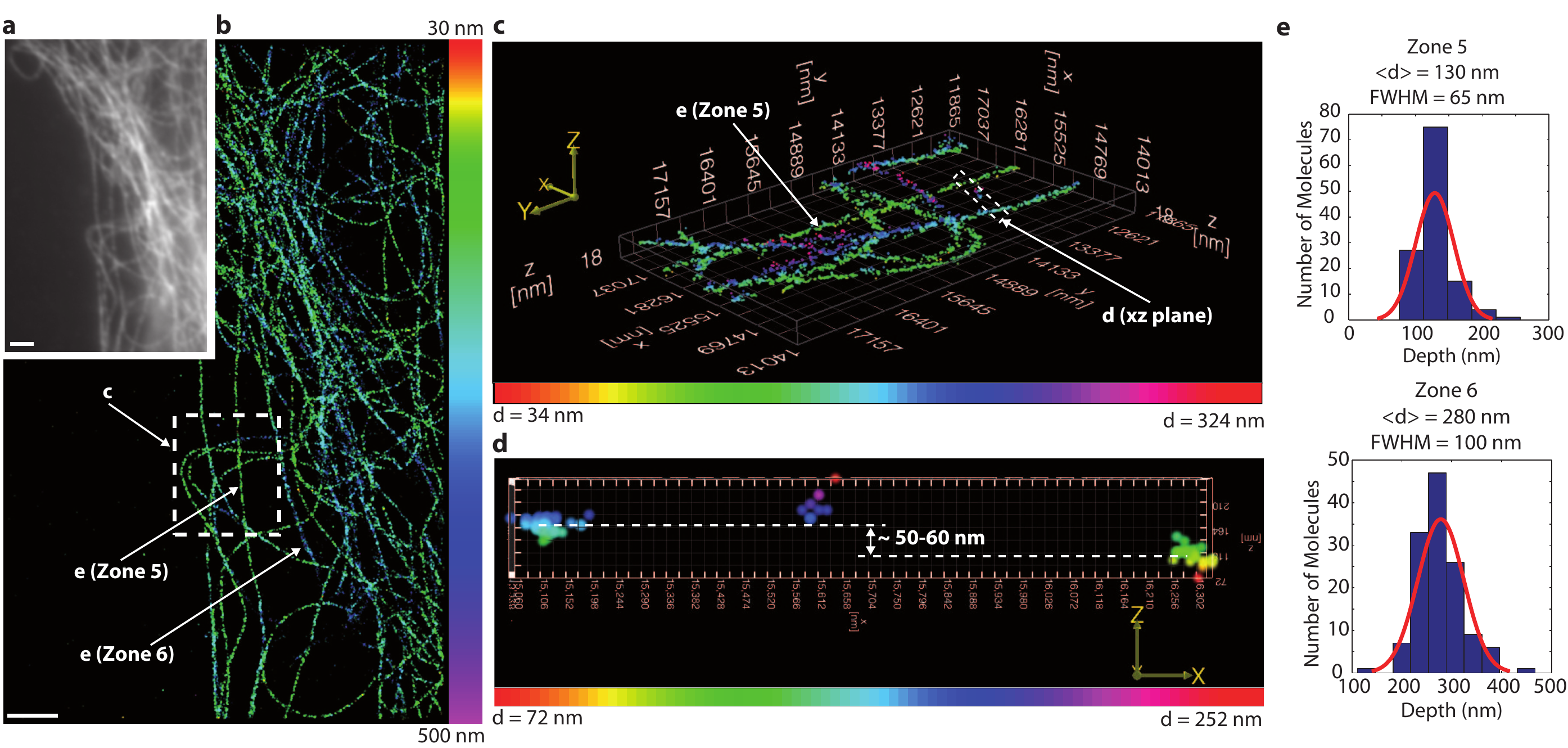} 
\caption{\textbf{dSTORM imaging of microtubules immersed in a Vectashield based buffer using DONALD. a} Conventional image of the microtubule network of immunofluorescently labeled CHO cells. \textbf{b} Three-dimensional super-resolved image of the same region shown in (\textbf{a}), color-coded as indicated by the colored depth-scale bar. \textbf{c} 3D visualization of the sub-region indicated in (\textbf{b}) by the dashed white box, produced using ViSP-software. \textbf{d} XZ-plane of the white-boxed region of (\textbf{c}), in which two microtubules are axially located with a depth-difference position of approximately 50-60 nm. \textbf{e} Axial profiles of two different microtubules (Zones 1 and 2) shown in (\textbf{b}) and (\textbf{c}). Axial resolution of 65 nm and 100 nm were obtained for filaments located at depths of approximately 130~nm and 280~nm, respectively. Scale bar: 2 $\mu m$ (\textbf{a,b})} 
\end{figure}
\end{spacing}
\end{document}